# A Dynamical Model of Harmonic Generation in Centrosymmetric Semiconductors

M. Scalora[1], M.A. Vincenti[2], D. de Ceglia[2], N. Akozbek[2], V. Roppo[3], M.J. Bloemer[1], J.W. Haus[4]

[1] Charles M. Bowden Research Center AMSRD-AMR-WSS, RDECOM, Redstone Arsenal, Alabama 35898-5000

[2] AEgis Technologies Group, 410 Jan Davis Dr., Huntsville, AL 35806

[3] Universitat Politècnica de Catalunya, Departament de Física i Eng. Nuclear, Colom 11, E-08222 Terrassa, Spain

[4] Electro-Optics Program, University of Dayton, 300 College Park, Dayton, OH 45469

## Abstract

We study second and third harmonic generation in centrosymmetric semiconductors at visible and UV wavelengths in bulk and cavity environments. Second harmonic generation is due to a combination of symmetry breaking, the magnetic portion of the Lorentz force, and quadrupolar contributions that impart peculiar features to the angular dependence of the generated signals, in analogy to what occurs in metals. The material is assumed to have a non-zero, third order nonlinearity that gives rise to most of the third harmonic signal. Using the parameters of bulk Silicon we predict that cavity environments can significantly modify second harmonic generation (390nm) with dramatic improvements for third harmonic generation (266nm). This occurs despite the fact that the harmonics may be tuned to a wavelength range where the dielectric function of the material is negative: a phase locking mechanism binds the pump to the generated signals and inhibits their absorption. These results point the way to novel uses and flexibility of materials like Silicon as nonlinear media in the visible and UV ranges.

PACS:

**Introduction**

Results on second harmonic generation (SHG) in centrosymmetric semiconductors like Silicon were first reported in reference [1]. A comparison of SHG between bulk Silicon and Silver showed that the angular dependence and the main features of the reflected signals were similar. These observations also suggested that SHG in Silicon was independent of angle cut and crystal orientation. The subject remained latent until the early 1980's, when new experimental results [2] implied that under intense illumination Silicon turned into a more ordinary material with a second order nonlinear coefficient having bulk dipolar origins. The authors argued that neither bulk quadrupolar sources nor symmetry breaking, impurities or superficial oxide layers were strong enough to explain their observations, and that the SH signal was consistent with a bulk-like $\chi^{(2)}$ similar to that of GaAs or GaP in symmetry and strength. Further experimental results [3] then suggested that the dependence of SHG on the azimuthal angle arose even at low intensities, and that surface and bulk sources gave way to comparable contributions. In light of these conflicting interpretations it was postulated that the experimental results could be explained by a combination of dipolar surface and quadrupolar bulk sources [4]. In a lengthy and detailed article the authors of reference [2] reiterated the conclusions expressed in their earlier work, that their experimental results could be explained only by the presence of bulk dipolar sources [5].

In the intervening years since this debate first took place a consensus has developed among researchers that materials like Si are characterized by intrinsic nonlinear surface anisotropy with surface dipolar and bulk quadrupolar sources [6-23] divided in almost equal parts. In reference [23] it was noted that even though bulk contributions are slightly smaller, the interference between these two types of sources cannot generally be neglected, making it difficult to identify and separate surface and bulk sources unambiguously [24, 25]. With recent advances in understanding of linear and nonlinear optical phenomena at the nanoscale many researchers are now taking a new look at well–known materials like Silicon and exploring the possibility of using them in new regimes and in novel ways. A recent example is provided by the investigation of harmonic generation in Silicon nanostructure at visible wavelengths [26].

As extensively reported in the literature, materials like Silicon are complicated by aspects related to nonlinear surface anisotropies [1-23]. Here we wish to build a self-consistent, classical model to describe frequency conversion and other nonlinear processes in a centrosymmetric, Silicon-like material that lacks a bulk $\chi^{(2)}$, may have a non-zero $\chi^{(3)}$, and where a combination of yet-to-be-determined surface and bulk sources gives rise to harmonic generation. Another aspect that we wish to underscore is that at visible wavelengths Silicon is highly absorptive, becoming metallic ($\mathrm{Re}(\varepsilon) < 0$) below 300nm [27]. We will show that a phase locking mechanism [28, 29] takes hold even in centrosymmetric media, rendering the material transparent to the harmonic wavelengths despite material dispersion, and that an exponential increase of THG occurs in the 80-300nm range in a Fabry-Perot etalon sandwiched between distributed Bragg reflectors (DBRs). The phase locked component corresponds to the inhomogeneous solution of the wave equation. In reference [29] it was experimentally demonstrated that GaP supports the propagation of phase locked third harmonic (TH) pulses at 223nm, in the metallic region.

A study of THG at infrared and visible wavelengths in absorbing materials including Si, Ge, and the noble metals [30] revealed the form of the $\chi^{(3)}$ tensor along with peculiar, dispersive properties due to differences that arise between conduction and valence electrons. A full band structure calculation of the dispersion of the $\chi^{(3)}$ tensor in Si, Ge, and GaAs was performed in reference [31]. As remarked in reference [32], bulk cubic materials always have a degree of anisotropy.  In isotropic media SH efficiency approaches zero when the pump field is incident normally on the surface [1], with different outcomes for TE or TM field polarizations, as shown for Ge (111) [33] and for particular crystallographic directions even for noble metals [34].

**The Model**

It is well-known that in noble metals bound charges contribute to SHG [35]. The role that bound charges play in SHG and THG was recently outlined in details using a harmonic oscillator model [36] and later used to describe harmonic generation in GaAs-filled, metallic nanocavities [37] and a GaP substrate [29]. In addition to being a powerful pedagogical tool, the nonlinear oscillator model provides a remarkably detailed description of dynamical factors that contribute to harmonic generation. The effects that arise from bound charges alone can be seen explicitly in the case of SHG and THG from a GaP substrate at UV wavelengths [29]. In what follows we generalize the model to include contributions of bound quadrupoles because in the absence of bulk nonlinearities (i.e. $\chi^{(2)}$, $\chi^{(3)}$, etc.) they can add significantly to the harmonic generation process. We will not consider the treatment of anisotropic surface sources, although the model can be easily modified to include arbitrary nonlinear surface distributions.

The description of a classical model usually begins with an equation of motion for a collection of uniformly distributed charges assumed to be under the action of internal forces (damping and harmonic restoring forces) and external forces due to the applied fields. A possible, quite simplistic pictorial representation of dipoles and quadrupoles present in the charge distribution is shown in Fig.1.  For an isotropic medium the equation of motion for the dipoles, described as a set of Lorentz oscillators, may be written as usual:

$$m_e^* \ddot{\mathbf{r}}_b + \gamma_e m_e^* \dot{\mathbf{r}}_b + k_e \mathbf{r}_b = e\mathbf{E} + \frac{e}{c}\dot{\mathbf{r}}_b \times \mathbf{B} \quad , \tag{1}$$

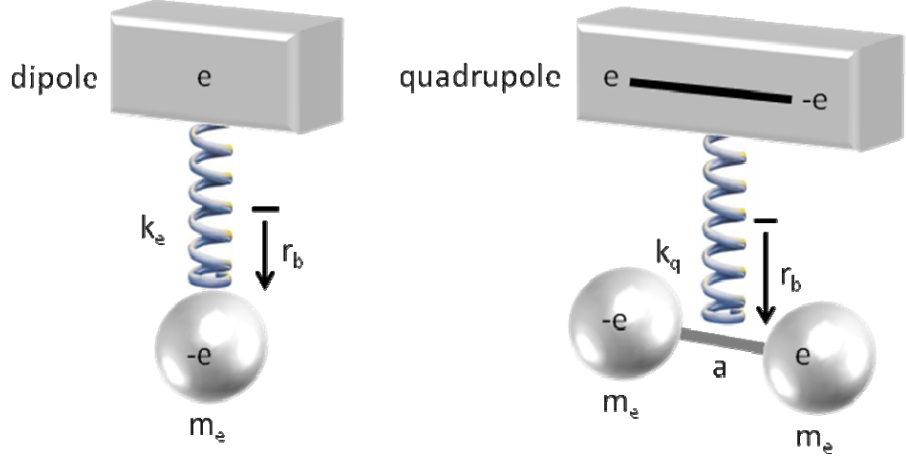

**Fig.1**: Depiction of classical dipole and quadrupole oscillators. One may assume that the distance $a$ between the two charges that make up the quadrupole is fixed, as suggested in Ref.[39].

where $m_e^*$ is the effective electron mass; $k_e$ is the spring constant; $e$ is the electron charge; $c$ is the speed of light in vacuum; $\mathbf{E}$ and $\mathbf{B(=H)}$ are the electric and magnetic fields at the electron's position; $\gamma_e$ is the damping coefficient; $\mathbf{r}_b$ is the bound electron's displacement from equilibrium.

An equation of motion may be written for the quadrupoles in similar fashion [38-40]:

$$m_q^*\ddot{\mathbf{r}}_b + \gamma_q m_q^*\dot{\mathbf{r}}_b + k_q\mathbf{r}_b = \frac{1}{2}\nabla\left(\sum_{j,k} Q_{jk} \frac{\partial E_j}{\partial x_k}\right) \qquad . \qquad (2)$$

where $Q_{jk}$ are the matrix elements of the quadrupole moment,

$$\mathbf{Q} = -\frac{e}{2}\mathbf{r}_b\mathbf{r}_b \ , \qquad (3)$$

$$U = -\frac{1}{2}\sum_{j,k} Q_{jk} \frac{\partial E_j}{\partial r_k} \qquad (4)$$

is the quadrupole-field interaction energy, and

$$\mathbf{F}_q = -\nabla U = \frac{1}{2}\nabla\left(\sum_{j,k} Q_{jk} \frac{\partial E_j}{\partial r_k}\right) \qquad (5)$$

is the net force experienced by the quadrupoles [39]. For simplicity we have assumed that dipoles and quadrupoles are co-located and share the same coordinate system, as indicated on Fig.1. Eqs.(1-2) then lead to a single, simplified equation of motion:

$$\ddot{\mathbf{r}}_b + \gamma_{eff}\dot{\mathbf{r}}_b + \omega_{eff}^2 \mathbf{r}_b = \frac{e}{m^*}\mathbf{E} + \frac{e}{m^*c}\dot{\mathbf{r}}_b \times \mathbf{B} + \frac{e}{4m^*}\nabla\left(\mathbf{r}_b\mathbf{r}_b : \nabla\mathbf{E}\right),$$ (6)

where we have assumed a simplified damping term relative to the corresponding dipolar term [39], and that $m^* = \left(m_e^* + m_q^*\right)$, $\gamma_{eff} = \left(\gamma_e m_e^* + \gamma_q m_q^*\right)/m^*$, and $\omega_{eff}^2 = \left(k_e + k_q\right)/m^*$.

We examine Eq.(6) by first expanding the quadrupolar terms explicitly in Cartesian coordinates:

$$\nabla\mathbf{E} = \begin{pmatrix} \dfrac{\partial}{\partial x}\mathbf{i} \\[2mm] \dfrac{\partial}{\partial y}\mathbf{j} \\[2mm] \dfrac{\partial}{\partial z}\mathbf{k} \end{pmatrix}\left(E_x\mathbf{i} + E_y\mathbf{j} + E_z\mathbf{k}\right) = \begin{pmatrix} \dfrac{\partial E_x}{\partial x} & \dfrac{\partial E_y}{\partial x} & \dfrac{\partial E_z}{\partial x} \\[2mm] \dfrac{\partial E_x}{\partial y} & \dfrac{\partial E_y}{\partial y} & \dfrac{\partial E_z}{\partial y} \\[2mm] \dfrac{\partial E_x}{\partial z} & \dfrac{\partial E_y}{\partial z} & \dfrac{\partial E_z}{\partial z} \end{pmatrix},$$ (7)

and

$$\mathbf{r}_b\mathbf{r}_b = \begin{pmatrix} x\mathbf{i} \\ y\mathbf{j} \\ z\mathbf{k} \end{pmatrix}\left(x\mathbf{i} \quad y\mathbf{j} \quad z\mathbf{k}\right) = \begin{pmatrix} x^2 & xy & xz \\ yx & y^2 & yz \\ zx & zy & z^2 \end{pmatrix}.$$ (8)

Combining Eqs.(7) and (8) the quadrupolar force is:

$$\nabla\left(\mathbf{r}_b\mathbf{r}_b : \nabla\mathbf{E}\right) = \left(\frac{\partial}{\partial x}\mathbf{i} + \frac{\partial}{\partial y}\mathbf{j} + \frac{\partial}{\partial z}\mathbf{k}\right)\begin{pmatrix} x^2\dfrac{\partial E_x}{\partial x} + xy\dfrac{\partial E_x}{\partial y} + xz\dfrac{\partial E_x}{\partial z} + yx\dfrac{\partial E_y}{\partial x} + y^2\dfrac{\partial E_y}{\partial y} \\[2mm] + yz\dfrac{\partial E_y}{\partial z} + zx\dfrac{\partial E_z}{\partial x} + zy\dfrac{\partial E_z}{\partial y} + z^2\dfrac{\partial E_z}{\partial z} \end{pmatrix}.$$ (9)

After a straightforward application of the spatial derivatives, Eq.(9) becomes:

$$\nabla\left(\mathbf{r}_b\mathbf{r}_b : \nabla\mathbf{E}\right) \approx \begin{pmatrix} \mathbf{i}\left(2x\dfrac{\partial E_x}{\partial x} + y\dfrac{\partial E_x}{\partial y} + z\dfrac{\partial E_x}{\partial z} + y\dfrac{\partial E_y}{\partial x} + z\dfrac{\partial E_z}{\partial x}\right) \\ \mathbf{j}\left(x\dfrac{\partial E_x}{\partial y} + x\dfrac{\partial E_y}{\partial x} + 2y\dfrac{\partial E_y}{\partial y} + z\dfrac{\partial E_y}{\partial z} + z\dfrac{\partial E_z}{\partial y}\right) \\ \mathbf{k}\left(x\dfrac{\partial E_x}{\partial z} + y\dfrac{\partial E_y}{\partial z} + x\dfrac{\partial E_z}{\partial x} + y\dfrac{\partial E_z}{\partial y} + 2z\dfrac{\partial E_z}{\partial z}\right) \end{pmatrix} \quad . \quad (10)$$

In writing Eq.(10) we have neglected terms that contain products of two coordinates and higher order derivatives of the fields (for example, $xy\dfrac{\partial^2 E_y}{\partial y^2}$) because particle displacements from equilibrium (i.e. $x$, $y$, and $z$) are intrinsically small. We can get another point of view if we cast Eq.(10) in terms of the macroscopic polarizations:

$$\nabla\left(\mathbf{r}_b\mathbf{r}_b : \nabla\mathbf{E}\right) \approx \frac{1}{ne}\begin{bmatrix} \mathbf{i}\left(2P_x\dfrac{\partial E_x}{\partial x} + P_y\dfrac{\partial E_x}{\partial y} + P_z\dfrac{\partial E_x}{\partial z} + P_y\dfrac{\partial E_y}{\partial x} + P_z\dfrac{\partial E_z}{\partial x}\right) \\ \mathbf{j}\left(P_x\dfrac{\partial E_x}{\partial y} + P_x\dfrac{\partial E_y}{\partial x} + 2P_y\dfrac{\partial E_y}{\partial y} + P_z\dfrac{\partial E_y}{\partial z} + P_z\dfrac{\partial E_z}{\partial y}\right) \\ \mathbf{k}\left(P_x\dfrac{\partial E_x}{\partial z} + P_y\dfrac{\partial E_y}{\partial z} + P_x\dfrac{\partial E_z}{\partial x} + P_y\dfrac{\partial E_z}{\partial y} + 2P_z\dfrac{\partial E_z}{\partial z}\right) \end{bmatrix}, \quad (11)$$

where we have identified the Cartesian components $P_j = ner_j$. We now simplify the geometry by assuming that all our structures have two-dimensional symmetry. With reference to Fig.2, we assume the wave propagates on the $y$-$z$ plane, and refer to $z$ as the longitudinal coordinate. Then, a TM-polarized field has E-field components that point along $y$ and $z$; a TE-polarized field has corresponding H-field components that point in the directions $y$ and $z$, and the fields are independent of $x$. Then, suppressing derivatives with respect to $x$, Eq.(11) simplifies to:

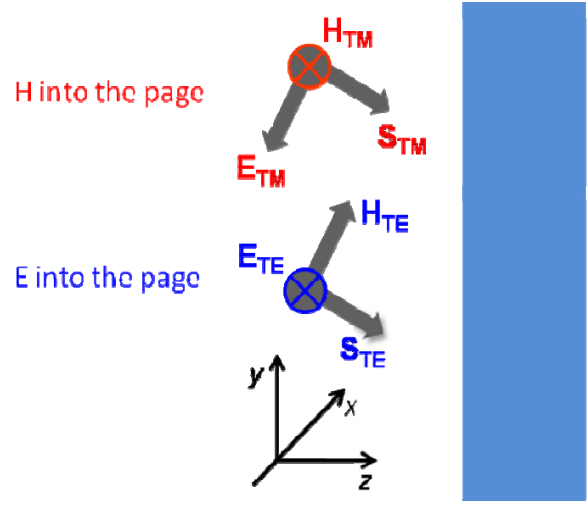

**Fig.2**: Typical geometry of a structure with two-dimensional symmetry. Propagation occurs on the plane y-z plane, and derivatives with respect to the $x$ are suppressed. The rectangular box represents the Si sample.

$$\nabla\left(\mathbf{r}_b\mathbf{r}_b:\nabla\mathbf{E}\right)\approx\frac{1}{ne}\begin{bmatrix}\mathbf{i}\left(P_y\dfrac{\partial E_x}{\partial y}+P_z\dfrac{\partial E_x}{\partial z}\right)\\[2mm]+\mathbf{j}\left(P_x\dfrac{\partial E_x}{\partial y}+2P_y\dfrac{\partial E_y}{\partial y}+P_z\dfrac{\partial E_y}{\partial z}+P_z\dfrac{\partial E_z}{\partial y}\right)\\[2mm]+\mathbf{k}\left(P_x\dfrac{\partial E_x}{\partial z}+P_y\dfrac{\partial E_y}{\partial z}+P_y\dfrac{\partial E_z}{\partial y}+2P_z\dfrac{\partial E_z}{\partial z}\right)\end{bmatrix}. \tag{12}$$

We now exploit the most salient points of the model presented in reference [36]. Assuming the presence of a fundamental field and two of its harmonics, we write the fields in terms of generic space and time dependent envelope functions, with carrier wave vector and frequency as follows:

$$\begin{aligned}\mathbf{E}&=E_x\mathbf{i}+E_y\mathbf{j}+E_z\mathbf{k}=\mathbf{E}_\omega e^{i(\mathbf{k}\bullet\mathbf{r}_b-\omega t)}+\mathbf{E}_{2\omega}e^{2i(\mathbf{k}\bullet\mathbf{r}_b-\omega t)}+\mathbf{E}_{3\omega}e^{3i(\mathbf{k}\bullet\mathbf{r}_b-\omega t)}+c.c.\\\mathbf{B}&=B_x\mathbf{i}+B_y\mathbf{j}+B_z\mathbf{k}=\mathbf{B}_\omega e^{i(\mathbf{k}\bullet\mathbf{r}_b-\omega t)}+\mathbf{B}_{2\omega}e^{2i(\mathbf{k}\bullet\mathbf{r}_b-\omega t)}+\mathbf{B}_{3\omega}e^{3i(\mathbf{k}\bullet\mathbf{r}_b-\omega t)}+c.c.\end{aligned}. \tag{13}$$

We expand the spatial portion of the exponential as a function of $\mathbf{k}\bullet\mathbf{r}_b$ up to second order:

$$\begin{aligned}\mathbf{E}=&\mathbf{E}_\omega e^{-i\omega t}+\mathbf{E}_\omega^*e^{i\omega t}+\mathbf{E}_{2\omega}e^{-2i\omega t}+\mathbf{E}_{2\omega}^*e^{2i\omega t}+\mathbf{E}_{3\omega}e^{-3i\omega t}+\mathbf{E}_{3\omega}^*e^{3i\omega t}\\&\left[\left(\mathbf{E}_\omega e^{-i\omega t}-\mathbf{E}_\omega^*e^{i\omega t}\right)+2\left(\mathbf{E}_{2\omega}e^{-2i\omega t}-\mathbf{E}_{2\omega}^*e^{2i\omega t}\right)+3\left(\mathbf{E}_{3\omega}e^{-3i\omega t}-\mathbf{E}_{3\omega}^*e^{3i\omega t}\right)\right]i\mathbf{k}\bullet\mathbf{r}_b\\&-\left[\left(\mathbf{E}_\omega e^{-i\omega t}+\mathbf{E}_\omega^*e^{i\omega t}\right)+4\left(\mathbf{E}_\omega e^{-2i\omega t}+\mathbf{E}_\omega^*e^{2i\omega t}\right)+9\left(\mathbf{E}_{3\omega}e^{-3i\omega t}+\mathbf{E}_{3\omega}^*e^{3i\omega t}\right)\right]\left(\mathbf{k}\bullet\mathbf{r}_b\right)^2/2\end{aligned}, \tag{14}$$

and similarly for **B**. Assuming a solution that once again involves generic envelope functions:

$$\mathbf{r}_b = \mathbf{r}_\omega \, e^{-i\omega t} + \mathbf{r}_{2\omega} \, e^{-2i\omega t} + \mathbf{r}_{3\omega} \, e^{-3i\omega t} + c.c. \qquad , \qquad (15)$$

substituting Eqs.(13-15) into Eq.(6), recognizing

$$\mathbf{P}_{b,\omega} = n_{0,b} e \mathbf{r}_\omega \qquad\qquad \mathbf{P}_{b,2\omega} = n_{0,b} e \mathbf{r}_{2\omega} \qquad\qquad \mathbf{P}_{b,3\omega} = n_{0,b} e \mathbf{r}_{3\omega} \qquad , \qquad (16)$$

approximating

$$i\mathbf{k} \bullet n_{0,b} e \mathbf{r}_\omega \approx \nabla \bullet \mathbf{P}_{b,\omega} \qquad 2i\mathbf{k} \bullet n_{0,b} e \mathbf{r}_{2\omega} \approx \nabla \bullet \mathbf{P}_{b,2\omega} \qquad 3i\mathbf{k} \bullet n_{0,b} e \mathbf{r}_{3\omega} \approx \nabla \bullet \mathbf{P}_{b,3\omega} \;, \qquad (17)$$

expanding Eq.(12) into its harmonic components and retaining lowest order terms finally leads to a set of three scaled, coupled equations similar to those found in reference [36]:

$$\begin{pmatrix} \ddot{\mathbf{P}}_\omega \\ +\tilde{\gamma}_\omega \dot{\mathbf{P}}_\omega \\ +\tilde{\omega}_{0,\omega}^2 \tilde{\mathbf{P}}_\omega \end{pmatrix} = \frac{e\lambda_0}{m_b^* c^2} \begin{pmatrix} ne\lambda_0 \mathbf{E}_\omega - \frac{1}{2}\left(\nabla \bullet \mathbf{P}_{2\omega}\right)\mathbf{E}_\omega^* + 2\left(\nabla \bullet \mathbf{P}_\omega^*\right)\mathbf{E}_{2\omega} \\ \frac{1}{ne\lambda_0}\left[ \mid \nabla \bullet \mathbf{P}_\omega \mid^2 \mathbf{E}_\omega - \frac{1}{2}\left(\nabla \bullet \mathbf{P}_\omega\right)^2 \mathbf{E}_\omega^* \right] \end{pmatrix} + \frac{e\lambda_0}{m_b^* c^2}\begin{pmatrix} \left(\dot{\mathbf{P}}_\omega^* + i\omega \mathbf{P}_\omega^*\right)\times \mathbf{H}_{2\omega} \\ +\left(\dot{\mathbf{P}}_{2\omega} - 2i\omega \mathbf{P}_{2\omega}\right)\times \mathbf{H}_\omega^* \end{pmatrix} + \frac{e\lambda_0}{4m_b^* c^2}\mathbf{F}_\omega$$

$$\begin{pmatrix} \ddot{\mathbf{P}}_{2\omega} \\ +\tilde{\gamma}_{2\omega}\dot{\mathbf{P}}_{2\omega} \\ +\tilde{\omega}_{0,2\omega}^2 \tilde{\mathbf{P}}_{2\omega} \end{pmatrix} = \frac{e\lambda_0}{m_b^* c^2}\begin{pmatrix} ne\lambda_0 \mathbf{E}_{2\omega} + \left(\nabla \bullet \mathbf{P}_\omega\right)\mathbf{E}_\omega - \frac{1}{3}\left(\nabla \bullet \mathbf{P}_{3\omega}\right)\mathbf{E}_\omega^* - 3\left(\nabla \bullet \mathbf{P}_\omega^*\right)\mathbf{E}_{3\omega} \\ \frac{1}{2ne\lambda_0}\begin{bmatrix} -\frac{1}{2}\left(\nabla \bullet \mathbf{P}_{2\omega}\right)\left(\nabla \bullet \mathbf{P}_\omega^*\right)\mathbf{E}_\omega \\ +\frac{1}{2}\left(\nabla \bullet \mathbf{P}_{2\omega}\right)\left(\nabla \bullet \mathbf{P}_\omega\right)\mathbf{E}_\omega^* - 2\mid\nabla \bullet \mathbf{P}_\omega\mid^2 \mathbf{E}_{2\omega} \end{bmatrix} \end{pmatrix} + \frac{e\lambda_0}{m_b^* c^2}\left(\dot{\mathbf{P}}_\omega - i\omega \mathbf{P}_\omega\right)\times \mathbf{H}_\omega + \frac{e\lambda_0}{4m_b^* c^2}\mathbf{F}_{2\omega}$$

$$\begin{pmatrix} \ddot{\mathbf{P}}_{3\omega} \\ +\tilde{\gamma}_{3\omega}\dot{\mathbf{P}}_{3\omega} \\ +\tilde{\omega}_{0,3\omega}^2 \mathbf{P}_{3\omega} \end{pmatrix} = \frac{e\lambda_0}{m_b^* c^2}\begin{pmatrix} ne\lambda_0 \mathbf{E}_{3\omega} + \frac{1}{2}\left(\nabla \bullet \mathbf{P}_{2\omega}\right)\mathbf{E}_\omega + 2\left(\nabla \bullet \mathbf{P}_\omega\right)\mathbf{E}_{2\omega} \\ +\frac{1}{2ne\lambda_0}\left(\nabla \bullet \mathbf{P}_\omega\right)^2 \mathbf{E}_\omega \end{pmatrix} + \frac{e\lambda_0}{m_b^* c^2}\begin{pmatrix} \left(\dot{\mathbf{P}}_{2\omega} - i\omega \mathbf{P}_{2\omega}\right)\times \mathbf{H}_\omega \\ +\left(\dot{\mathbf{P}}_\omega - i\omega \mathbf{P}_\omega\right)\times \mathbf{H}_{2\omega} \\ +\frac{1}{ne\lambda_0}\left(\dot{\mathbf{P}}_\omega - i\omega \mathbf{P}_\omega\right)\times \mathbf{H}_\omega \nabla \bullet \mathbf{P}_\omega \end{pmatrix} + \frac{e\lambda_0}{4m_b^* c^2}\mathbf{F}_{3\omega}$$

$$\qquad , \qquad (18)$$

where

$$\mathbf{F}_\omega = \mathbf{i}\left(\left(\frac{\partial E_{x,2\omega}}{\partial y} + 2ik_y E_{x,2\omega}\right)P_{y,\omega}^* + \left(\frac{\partial E_{x,\omega}}{\partial y} + ik_y E_{x,\omega}\right)^* P_{y,2\omega} + \left(\frac{\partial E_{x,2\omega}}{\partial z} + 2ik_z E_{x,2\omega}\right)P_{z,\omega}^* + \left(\frac{\partial E_{x,\omega}}{\partial z} + ik_z E_{x,\omega}\right)^* P_{z,2\omega}\right)$$

$$+\mathbf{j}\left(\begin{array}{l}\left(\frac{\partial E_{x,2\omega}}{\partial y} + 2ik_y E_{x,2\omega}\right)P_{x,\omega}^* + \left(\frac{\partial E_{x,\omega}}{\partial y} + ik_y E_{x,\omega}\right)^* P_{x,2\omega} + 2\left(\frac{\partial E_{y,2\omega}}{\partial y} + 2ik_y E_{y,2\omega}\right)P_{y,\omega}^* + 2\left(\frac{\partial E_{y,\omega}}{\partial y} + ik_y E_{y,\omega}\right)^* P_{y,2\omega} \\ + \left(\frac{\partial E_{y,2\omega}}{\partial z} + 2ik_z E_{y,2\omega}\right)P_{z,\omega}^* + \left(\frac{\partial E_{z,2\omega}}{\partial y} + 2ik_y E_{z,2\omega}\right)P_{z,\omega}^* + \left(\frac{\partial E_{y,\omega}}{\partial z} + ik_z E_{y,\omega}\right)^* P_{z,2\omega} + \left(\frac{\partial E_{z,\omega}}{\partial y} + ik_y E_{z,\omega}\right)^* P_{z,2\omega}\end{array}\right)$$

$$+\mathbf{k}\left(\begin{array}{l}2\left(\frac{\partial E_{z,2\omega}}{\partial z} + 2ik_z E_{z,2\omega}\right)P_{z,\omega}^* + 2\left(\frac{\partial E_{z,\omega}}{\partial z} + ik_z E_{z,\omega}\right)^* P_{z,2\omega} + \left(\frac{\partial E_{y,2\omega}}{\partial z} + 2ik_z E_{y,2\omega}\right)P_{y,\omega}^* + \left(\frac{\partial E_{z,2\omega}}{\partial y} + 2ik_y E_{z,2\omega}\right)P_{y,\omega}^* \\ + \left(\frac{\partial E_{y,\omega}}{\partial z} + ik_z E_{y,\omega}\right)^* P_{y,2\omega} + \left(\frac{\partial E_{z,\omega}}{\partial y} + ik_y E_{z,\omega}\right)^* P_{y,2\omega} + \left(\frac{\partial E_{x,2\omega}}{\partial z} + 2ik_z E_{x,2\omega}\right)P_{x,\omega}^* + \left(\frac{\partial E_{x,\omega}}{\partial z} + ik_z E_{x,\omega}\right)^* P_{x,2\omega}\end{array}\right)$$

$$, \qquad (19)$$

$$\mathbf{F}_{2\omega} = \mathbf{i}\left(\left(\frac{\partial E_{x,\omega}}{\partial y} + ik_y E_{x,\omega}\right)P_{y,\omega} + \left(\frac{\partial E_{x,\omega}}{\partial z} + ik_z E_{x,\omega}\right)P_{z,\omega}\right)$$

$$+\mathbf{j}\left(\left(\frac{\partial E_{x,\omega}}{\partial y} + ik_y E_{x,\omega}\right)P_{x,\omega} + 2\left(\frac{\partial E_{y,\omega}}{\partial y} + ik_y E_{y,\omega}\right)P_{y,\omega} + \left(\frac{\partial E_{y,\omega}}{\partial z} + ik_z E_{y,\omega} + \frac{\partial E_{z,\omega}}{\partial y} + ik_y E_{z,\omega}\right)P_{z,\omega}\right) \qquad , \qquad (20)$$

$$+\mathbf{k}\left(\left(\frac{\partial E_{x,\omega}}{\partial z} + ik_z E_{x,\omega}\right)P_{x,\omega} + 2\left(\frac{\partial E_{z,\omega}}{\partial z} + ik_z E_{z,\omega}\right)P_{z,\omega} + \left(\frac{\partial E_{y,\omega}}{\partial z} + ik_z E_{y,\omega} + \frac{\partial E_{z,\omega}}{\partial y} + ik_y E_{z,\omega}\right)P_{y,\omega}\right)$$

and

$$\mathbf{F}_{3\omega} = \mathbf{i}\left(\left(\frac{\partial E_{x,2\omega}}{\partial y} + 2ik_y E_{x,2\omega}\right)P_{y,\omega} + \left(\frac{\partial E_{x,\omega}}{\partial y} + ik_y E_{x,\omega}\right)P_{y,2\omega} + \left(\frac{\partial E_{x,2\omega}}{\partial z} + 2ik_z E_{x,2\omega}\right)P_{z,\omega} + \left(\frac{\partial E_{x,\omega}}{\partial z} + ik_z E_{x,\omega}\right)P_{z,2\omega}\right)$$

$$+\mathbf{j}\left(\begin{array}{l}\left(\frac{\partial E_{x,2\omega}}{\partial y} + 2ik_y E_{x,2\omega}\right)P_{x,\omega} + \left(\frac{\partial E_{x,\omega}}{\partial y} + ik_y E_{x,\omega}\right)P_{x,2\omega} + 2\left(\frac{\partial E_{y,2\omega}}{\partial y} + 2ik_y E_{y,2\omega}\right)P_{y,\omega} + 2\left(\frac{\partial E_{y,\omega}}{\partial y} + ik_y E_{y,\omega}\right)P_{y,2\omega} \\ + \left(\frac{\partial E_{z,2\omega}}{\partial y} + 2ik_z E_{z,2\omega} + \frac{\partial E_{y,2\omega}}{\partial z} + 2ik_y E_{y,2\omega}\right)P_{z,\omega} + \left(\frac{\partial E_{z,\omega}}{\partial y} + ik_z E_{z,\omega} + \frac{\partial E_{y,\omega}}{\partial z} + ik_z E_{y,\omega}\right)P_{z,2\omega}\end{array}\right)$$

$$+\mathbf{k}\left(\begin{array}{l}\left(\frac{\partial E_{x,2\omega}}{\partial z} + 2ik_z E_{x,2\omega}\right)P_{x,\omega} + \left(\frac{\partial E_{x,\omega}}{\partial z} + ik_z E_{x,\omega}\right)P_{x,2\omega} + \left(\frac{\partial E_{y,2\omega}}{\partial z} + 2ik_y E_{y,2\omega} + \frac{\partial E_{z,2\omega}}{\partial y} + 2ik_z E_{z,2\omega}\right)P_{y,\omega} \\ + \left(\frac{\partial E_{y,\omega}}{\partial z} + ik_z E_{y,\omega} + \frac{\partial E_{z,\omega}}{\partial y} + ik_y E_{z,\omega}\right)P_{y,2\omega} + 2\left(\frac{\partial E_{z,2\omega}}{\partial z} + 2ik_z E_{z,2\omega}\right)P_{z,\omega} + 2\left(\frac{\partial E_{z,\omega}}{\partial z} + ik_z E_{z,\omega}\right)P_{z,2\omega}\end{array}\right)$$

$$(21)$$

are the lowest order quadrupolar contributions. In writing Eqs.(19-21) we have ignored higher order quadrupolar terms like

$$\left(\frac{e\lambda_0}{4m_b^*c^2}\right)\left(\frac{1}{ne\lambda_0}\right)P_{y,\omega}\left(\frac{\partial E_{y,2\omega}}{\partial y}+2ik_yE_{y,2\omega}\right)\nabla\bullet\mathbf{P}_\omega^*e^{-2i\omega t}, \quad \text{because} \quad 1/ne\lambda_0\sim10^{-8}-10^{-10}\text{(cgs)} \quad \text{for}$$

materials that range from dielectrics to metals. Differentiation is now with respect to the dimensionless coordinates, $\xi=z/\lambda_0$, $\tilde{y}=y/\lambda_0$, and $\tau=ct/\lambda_0$, and we have chosen $\lambda_0=1\mu m$. Eqs.(18) are solved in the time domain together with Maxwell's equations using a fast Fourier transform pulse propagation method [36].

At this point some comments that relate to the numerical integration of Eqs.(18) are in order. First, the validity of the expansion in Eq.(14) may be called into question if the boundary between two media were infinitely sharp. In tackling any kind of numerical problem there are at least two important questions that one must always ask: Does the calculation converge? Is the result accurate? One of the more obvious and somewhat prosaic aspects of actually performing the calculation in the time domain is that it requires the use of finite spatial and temporal time steps, and pulses that always have a finite turn-on time. The polarization thus grows from zero as a function of time as the pulse approaches, and for different spatial step sizes at any instant the boundary may look like that depicted in Fig.3. The conjunction points at the top and bottom of the oblique sections of the curves could be smoothed out to insure that the functions are well-behaved and differentiable everywhere. It turns out that this operation is unnecessary. Usually, if the equations of motion do not contain intrinsic instabilities the calculation converges rapidly to a given trajectory. Convergence is checked by decreasing spatial and temporal step sizes simultaneously. While decreasing spatial step size leads to functions that display the geometry of Fig.3, a smaller time step causes the pulse to advance and to sample the boundary at slower rates. Then, the derivatives of typical variables can be large but they are always finite, and "hard" interfaces are in fact quite manageable provided enough terms are retained in the Taylor

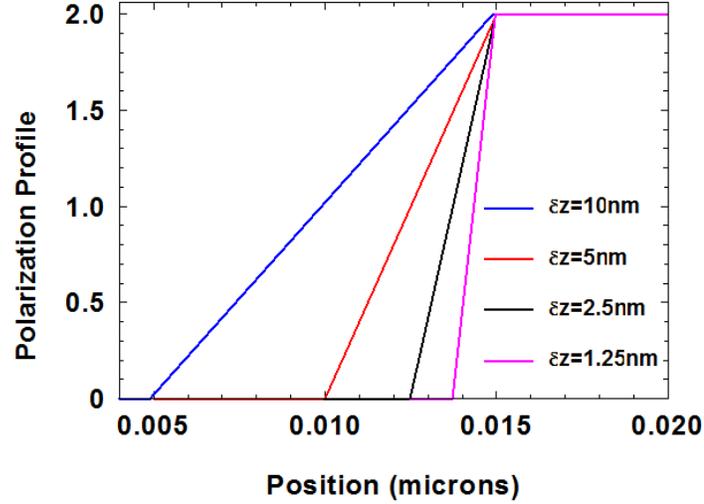

**Fig.3:** Typical polarization profile as a function of position for different spatial integration steps.

expansion in Eq.(14) around neighborhoods with steep slopes.

The other issue has to do with the kinds of terms that contribute to harmonic generation. Eqs.(18) imply that there are three types of sources: (i) terms proportional to $\nabla \bullet \mathbf{P}_{\omega}$ or $\nabla \bullet \mathbf{P}_{2\omega}$, for example, that are zero everywhere except at the surface, have quadrupolar character but dipolar origin, may trigger self-phase modulation for the pump, and contribute most to THG if $\chi^{(2)} = \chi^{(3)} = 0$; (ii) magnetic contributions like $\left( \dot{\mathbf{P}}_{\omega} - i\omega \mathbf{P}_{\omega} \right) \times \mathbf{H}_{\omega}$ and $\left( \dot{\mathbf{P}}_{2\omega} - i\omega \mathbf{P}_{2\omega} \right) \times \mathbf{H}_{\omega}$, which can account the dynamics of ultrashort pulses; (iii) quadrupolar contributions-Eq.(11)-that contain derivatives of the polarizations and fields, that exhibit surface and volume qualities simultaneously since we are dealing with pulses. We now provide some perspective on the quadrupolar terms in Eqs.(11-12) by comparing them with quadrupolar terms that are extracted when one uses energy arguments to define nonlinear polarization sources. For the case of SHG, nonlinear quadrupolar sources may be expressed as [1, 3, 4, 22, 24, 25, 38, 41, 43, 45]:

$$P_i^{2\omega} = \sum_{jkl} \chi_{ijkl}^{(2)} E_j \nabla_k E_l \qquad\qquad , \qquad\qquad (22)$$

where $i,j,k,l$ are Cartesian coordinates, and $\chi^{(2)}_{ijkl}$ is a tensor that has the same crystallographic properties of the medium. For cubic materials $\chi^{(2)}_{ijkl}$ has twenty one non-zero elements, four of which are independent [42]. Expanding Eq.(22) and neglecting derivatives with respect to $x$:

$$
\begin{pmatrix} P_x^{2\omega} \\ P_y^{2\omega} \\ P_z^{2\omega} \end{pmatrix} = \begin{pmatrix} \chi^{(2)}_{xyyx}E_y\dfrac{\partial E_x}{\partial y} + \chi^{(2)}_{xzzx}E_z\dfrac{\partial E_x}{\partial z} + \chi^{(2)}_{xxyy}E_x\dfrac{\partial E_y}{\partial y} + \chi^{(2)}_{xxzz}E_x\dfrac{\partial E_z}{\partial z} \\[2mm] \chi^{(2)}_{yxyx}E_x\dfrac{\partial E_x}{\partial y} + \chi^{(2)}_{yyyy}E_y\dfrac{\partial E_y}{\partial y} + \chi^{(2)}_{yzzy}E_z\dfrac{\partial E_y}{\partial z} + \chi^{(2)}_{yzyz}E_z\dfrac{\partial E_z}{\partial y} + \chi^{(2)}_{yyzz}E_y\dfrac{\partial E_z}{\partial z} \\[2mm] \chi^{(2)}_{zxzx}E_x\dfrac{\partial E_x}{\partial z} + \chi^{(2)}_{zzzy}E_z\dfrac{\partial E_y}{\partial y} + \chi^{(2)}_{zzyz}E_y\dfrac{\partial E_y}{\partial z} + \chi^{(2)}_{zyyz}E_y\dfrac{\partial E_z}{\partial y} + \chi^{(2)}_{zzzz}E_z\dfrac{\partial E_z}{\partial z} \end{pmatrix}.
$$
(23)

Although the polarization components $P_j$ are rather complicated solutions of Eqs.(18), for discussion purposes only we may arbitrarily extract the leading term and write it in the form $P_j = \chi_L E_j$, where $\chi_L$ is the linear susceptibility, so that Eq.(12) becomes:

$$
\nabla\left(\mathbf{r}_b\mathbf{r}_b : \nabla\mathbf{E}\right) \approx \frac{\chi_L}{ne}\begin{bmatrix} \mathbf{i}\left( E_y\dfrac{\partial E_x}{\partial y} + E_z\dfrac{\partial E_x}{\partial z} \right) \\[2mm] \mathbf{j}\left( E_x\dfrac{\partial E_x}{\partial y} + 2E_y\dfrac{\partial E_y}{\partial y} + E_z\dfrac{\partial E_y}{\partial z} + E_z\dfrac{\partial E_z}{\partial y} \right) \\[2mm] \mathbf{k}\left( E_x\dfrac{\partial E_x}{\partial z} + E_y\dfrac{\partial E_y}{\partial z} + E_y\dfrac{\partial E_z}{\partial y} + 2E_z\dfrac{\partial E_z}{\partial z} \right) \end{bmatrix}.
$$
(24)

A comparison between Eqs.(23-24) reveals that aside from the proportionality constants the majority of terms are identical. However, there is a disparity in the number of terms present for each component. While on one hand the model appears to capture the basic dynamics of the system despite its simplicity, on the other it cannot fully account for the crystalline nature of materials. This shortcoming may be cured by going back to Eq.(2) and introducing a generalized quadrupole moment that contains an anisotropic portion that allows it to couple with all possible field gradient projections within the interaction volume. The anisotropic portion, written as

$$\mathbf{Q}_{anisotropic} = -\frac{e}{2} \begin{pmatrix} \beta_{xx} y^2 + \delta_{xx} z^2 & \beta_{xy} zy + \delta_{xy} xz & \beta_{xz} xy + \delta_{xz} yz \\ \beta_{yx} yz + \delta_{yx} xz & \beta_{yy} x^2 + \delta_{yy} z^2 & \beta_{yz} xy + \delta_{yz} xz \\ \beta_{zx} xy + \delta_{zx} yz & \beta_{zy} xy + \delta_{zy} xz & \beta_{zz} y^2 + \delta_{zz} x^2 \end{pmatrix}, \tag{25}$$

where $\beta_{ij}$ and $\delta_{ij}$ are constants to be determined, may be added to Eq.(3). One may then repeat the procedure beginning with Eq.(4) and all terms in Eq.(23) are recovered. A generalized mathematical approach requires expanding Eq.(2) using all Cartesian components to account for the distribution of charge, but supplementing Eqs.(2-3) with Eq.(25) suffices for our purposes.

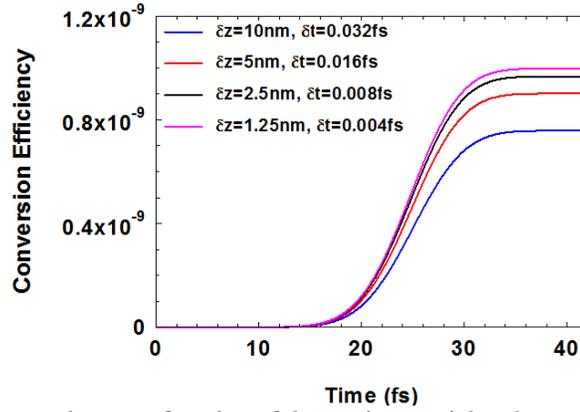

**Fig.4:** Total reflected efficiency vs. time as a function of decreasing spatial and temporal step sizes. The calculations clearly provide a fairly accurate answer even for the largest integration steps shown. In scaled units, $\delta\xi = \delta\tau$.

We now show an example that clearly demonstrates stability, convergence, and accuracy when Eqs.(18) are integrated together with Maxwell's equations using a quadrupole moment apt for cubic materials. We calculate reflected SH efficiencies from a thick Silicon substrate as a function of time by progressively decreasing spatial and temporal integration steps, and plot the results in Fig.4. TM-polarized incident pulses are approximately 20fs in duration, the carrier wavelength is λ=780nm, and the angle of incidence is 30º. The figure clearly shows that the calculations are stable even for relatively large integration steps, with converging solutions that become progressively more accurate with decreasing step size. Emitted field amplitudes saturate and further reductions of the integration steps beyond those shown in the figure add no discernable changes to the total efficiency or to the generated signals. Fig.4 demonstrates the validity of Eq.(14), that all spatial derivatives in Eqs.(18) are manageable, that one is not required to introduce *ad hoc* surface nonlinearities, or guess their signs and amplitudes because they are determined dynamically as the pulse crosses the interface. This makes the addition of anisotropic surface sources in a 3-D model relatively easy to implement.

Finally, we detail the inclusion of $\chi^{(3)}$ in our model [37]. Although part of the TH signal originates from terms in the last of Eqs.(18), a non-zero $\chi^{(3)}$ is responsible for most THG. To account for the additional terms we define a nonlinear polarization as usual:

$$P_i^{NL} = \sum_{j=1,3,} \sum_{k=1,3,} \sum_{l=1,3,} \chi_{ijkl}^{(3)} E_j E_k E_l \qquad , \tag{26}$$

where *i,j,k,l* are Cartesian coordinates. Eq.(26) yields a total of eighty one components [42], with non-zero terms that closely resemble those of Eq.(23). The result is:

$$P_x^{NL} = \chi_{xxxx}^{(3)} E_x^3 + \left( \chi_{xyyx}^{(3)} + \chi_{xyxy}^{(3)} + \chi_{xxyy}^{(3)} \right) E_y^2 E_x + \left( \chi_{xzzx}^{(3)} + \chi_{xzxz}^{(3)} + \chi_{xxzz}^{(3)} \right) E_z^2 E_x$$
$$P_y^{NL} = \chi_{yyyy}^{(3)} E_y^3 + \left( \chi_{yyxx}^{(3)} + \chi_{yxyx}^{(3)} + \chi_{yxxy}^{(3)} \right) E_x^2 E_y + \left( \chi_{yzzy}^{(3)} + \chi_{yzyz}^{(3)} + \chi_{yyzz}^{(3)} \right) E_z^2 E_y \qquad . \tag{27}$$
$$P_z^{NL} = \chi_{zzzz}^{(3)} E_z^3 + \left( \chi_{zzxx}^{(3)} + \chi_{zxzx}^{(3)} + \chi_{zxxz}^{(3)} \right) E_x^2 E_z + \left( \chi_{zzyy}^{(3)} + \chi_{zyzy}^{(3)} + \chi_{zyyz}^{(3)} \right) E_y^2 E_z$$

For simplicity we may set all coefficients equal, so that Eq.(27) may be simplified to read:

$$P_x^{NL} = \chi^{(3)} \left( E_x^2 + E_y^2 + E_z^2 \right) E_x$$
$$P_z^{NL} = \chi^{(3)} \left( E_x^2 + E_y^2 + E_z^2 \right) E_y \qquad . \tag{28}$$
$$P_z^{NL} = \chi^{(3)} \left( E_x^2 + E_y^2 + E_z^2 \right) E_z$$

Substitution of Eqs.(13) into Eq.(28) obviously generates nonlinear source terms for the pump and its harmonics, that are then inserted back into Eqs.(18). In what follows we choose $\chi_{cgs}^{(3)} = 0.75 \times 10^{-10}$ esu, equivalent to $\chi_{MKS}^{(3)} \sim 10^{-18} (\text{m/V})^2$.

**SHG and THG from bulk Silicon**

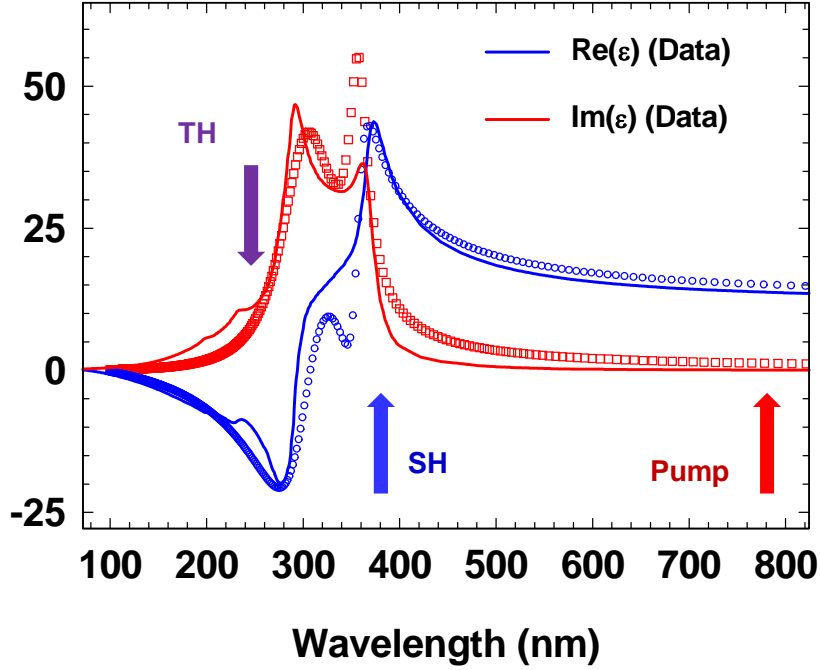

**Fig.5:** Dielectric response of Silicon at visible and near IR wavelengths. The thin solid curves correspond to the data, as indicated. Empty circles and squares that retrace the complex data set correspond to a plot of Eq.(28). The data is reproduced reasonably well across the entire range.

Now that we have tested numerical stability and convergence we examine harmonic generation from a thick Silicon substrate. We use the same quadrupole moment used to compute Fig.4, one that yields all terms displayed in Eqs.(23) with the relevant $\beta_{ij}$ and $\delta_{ij}$ in Eq.(25) set equal to unity, so that all terms are placed on equal footing. Then, the largest contributions will come from just a few terms proportional to $\partial E_z / \partial z$. The complex dielectric function of Silicon [27] is plotted in Fig.5. There are two absorption resonances in the UV range, and Re(ε)<0 for λ < 300nm. The linear data is fitted using two detuned Lorentz oscillators, so that:

$$\varepsilon(\omega) = 1 - \frac{\omega_{p,1}^2}{\omega^2 - \omega_{0,1}^2 + i\gamma_1 \omega} - \frac{\omega_{p,2}^2}{\omega^2 - \omega_{0,2}^2 + i\gamma_2 \omega};$$

(29)

$\omega$ is in units of $\mu m^{-1}$, $\gamma_1, \gamma_2 = 0.75, 0.18$ are damping constants, $\omega_{0,1}, \omega_{0,2} = 3.3, 2.8$ are resonance frequencies, and $\omega_{p,1}, \omega_{p,2} = 10, 4.5$ are plasma frequencies. Real and imaginary parts of $\varepsilon(\omega)$ are shown in Fig.5. We then integrate two sets of Eqs.(18) simultaneously, each corresponding to a different oscillator each having its own quadrupolar species. Incident pulses with peak intensities of 0.5GW/cm$^2$ are tuned to 780nm, where the absorption length is ~25$\mu$m. SH and TH signals are tuned to an absorption resonance and to the metallic range, respectively.

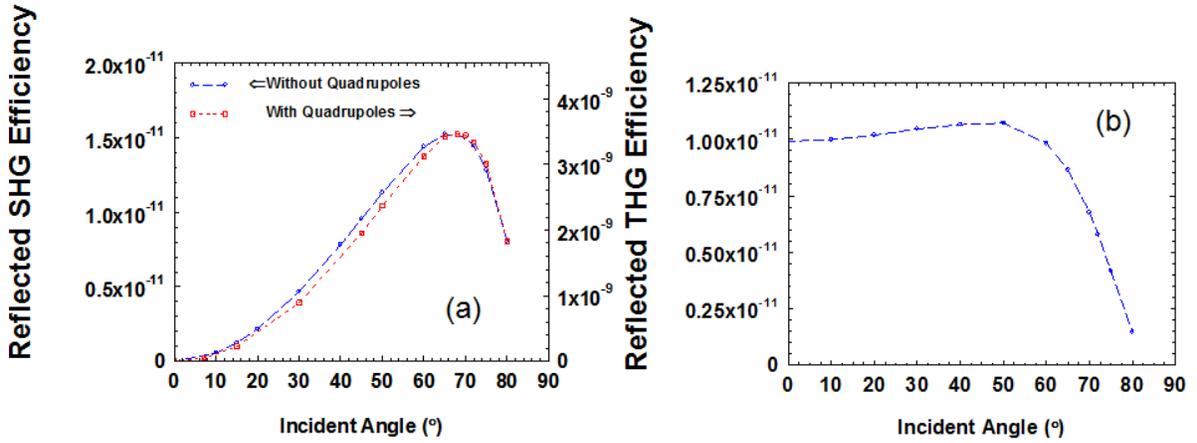

**Fig.6:** SHG (a) and THG (b) as a function of incident angle for a bulk Silicon substrate. The comparison in (a) suggests that most of the SH signal has quadrupolar origin. For THG, on the other hand (b), the signal originates mostly with the bulk $\chi^{(3)}$, and closely resembles the predictions in reference [30].

In Fig.6 we plot the calculated reflected SHG (a) and THG (b) conversion efficiencies with and without quadrupolar contributions. The SHG maxima are separated by just a few degrees, and with all else being equal the quadrupoles enhance the intensity of the emitted signal by more than two orders of magnitude (right axis scale) without changing the basic shape. If all matrix elements of the quadrupole moment are taken to have equal weight then most of the signal originates with the quadrupolar term $P_y^{2\omega} = \chi_{yyzz}^{(2)} E_y \left( \partial E_z / \partial z \right)$ in Eq.(23), or equivalently, with the more generic term $P_y^{2\omega} = P_y \left( \partial E_z / \partial z \right) / ne$ added to Eq.(12). In contrast, reflected THG is largely unaffected by sources unrelated to the $\chi^{(3)}$, and retains the same shape and amplitude regardless of the circumstance. In particular, we note that its shape is nearly identical to the Fresnel factor used to calculate TH conversion efficiencies from bulk Silicon in reference [30].

**SHG and THG from a Fabry-Perot Etalon and Multilayer Stack**

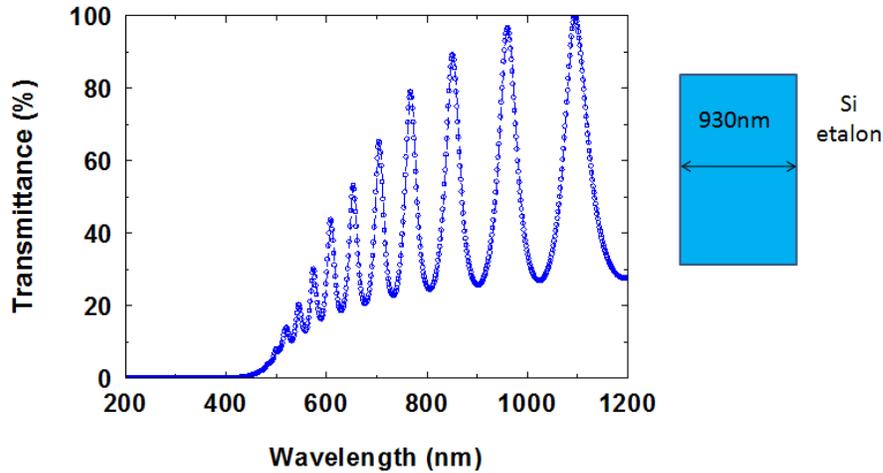

**Fig.7** Linear transmittance for a Silicon etalon 930nm thick at normal incidence. At 780nm transmittance is ~80%; wavelengths below 450nm are absorbed, with no expectation to see harmonic signals pass through the etalon.

While the calculations for a thick substrate suggest that the shape of the emitted SH signal remains the same with or without quadrupole contributions, this is no longer the case for a Fabry-Perot etalon. In Fig.7 we show the linear transmission function vs. incident wavelength for a Si layer 930nm thick, in the UV-Near IR range. Although a pump field tuned near 780nm is still resonantly transmitted at a rate of ~80%, the figure suggests that there is no expectation to see any SH (390nm) or TH (260nm) pass through the etalon: below 450nm transmittance is shut down by linear material absorption. Two mechanisms allow the observation of transmitted SH and/or TH signals: (i) phase locking, so that part of the signal that exits the etalon is forcibly carried by the pump, just as it occurs for a GaP substrate [29], and (ii) the interaction of the pump with the exit surface. Both contributions may be similar for zero $\chi^{(2)}$ and/or $\chi^{(3)}$. In this case the transmitted TH signal comes mostly from the dominant $\chi^{(3)}$ interaction.

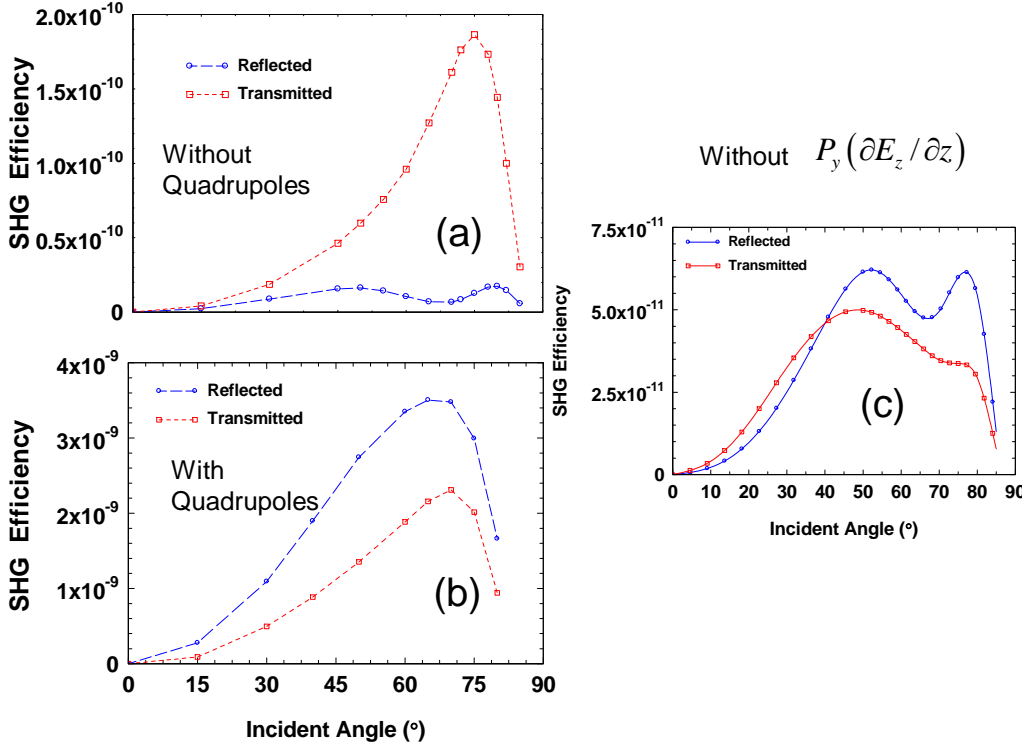

**Fig.8:** SHG without (a) and with (b) quadrupolar contributions, and (c) without the most dominant quadrupolar term. The shapes of the curves in 8(b) depend on the relative amplitudes of the matrix elements of the quadrupole moment. 8(c) shows that the curves take on configurations intermediate between 8(a) and 8(b), with efficiencies similar to 8(a), so that the types of nonlinear sources discussed in the text contribute to similar degrees.

In Figs.8 we plot SH efficiencies vs. incident angle with and without quadrupolar contributions for the etalon depicted in Fig.7. From Fig.8b one can see that quadrupoles can significantly remodulate the shape and the amplitude of the signal. A comparison between the conversion efficiencies of bulk (Fig.6a) and etalon (Fig.8b) shows that the reflected signal is practically unchanged, while the transmitted signal in Fig.8(b) is now quite substantial. The actual shapes of the transmitted and reflected curves shown will depend on the relative magnitudes of the matrix elements in Eqs.(23), and may ultimately be somewhere between the shapes of Figs.(8a) and (8b). An example of how the curves reshape is shown in Fig.8(c), where SHG efficiency is calculated without the most dominant quadrupolar term. These results show that any attempt to assign more importance to one type of source or another should probably be tied to more precise knowledge of the matrix elements in Eq.(23).

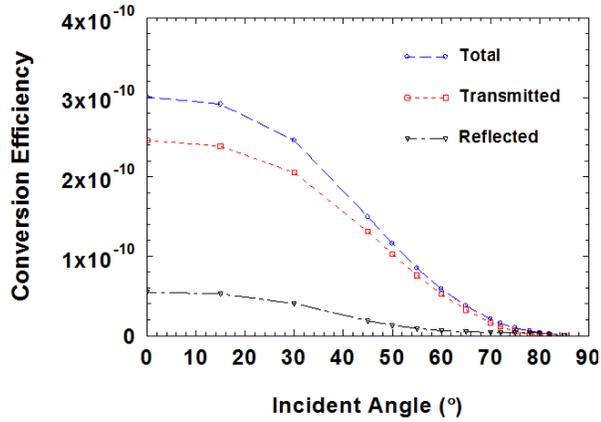

**Fig.9:** THG vs incident angle for the Fabry-Perot etalon depicted in Fig.(7). When this curve is compared with Fig.(6), which depicts THG from bulk silicon, one can deduce that total efficiency is amplified by a factor of ~30. Most of the enhancement comes from the transmitted signal, which is phase locked and resonates with the pump.

The results of THG for the etalon are far less ambiguous and are shown in Fig.9. The total THG efficiency for the etalon is enhanced by a factor of 30 relative to bulk (Fig.6b), most of it coming from the transmitted portion of the signal at normal incidence. As we mentioned above, one of the mechanisms that allows SH and TH signals to be transmitted in environments that are either strongly absorptive or metallic is phase locking: the pump traps the harmonics and imparts to them its own dispersive properties. It then becomes possible for the harmonics to resonate, even though the cavity is designed to resonate only at the fundamental wavelength [44]. THG has already been experimentally verified for bulk GaP in the metallic range [29], but there is still no experimental evidence that cavities can be used to enhance THG if the dielectric constant of the active layer is negative. Here we predict that the enhancement of THG can in fact be significant if we consider a Si etalon sandwiched between two DBRs. The multilayer stack - Fig.10- consists of a Si layer 330nm thick sandwiched between two, three-period $SiO_2$(100nm)/$Si_3N_4$(140nm) DBRs. Linear transmittance and reflectance are shown in Fig.10, with a Fabry-Perot resonance at ~813nm. The estimated cavity Q=$\lambda/\delta\lambda$~40. In Fig.11 we plot transmitted and reflected TH efficiencies vs. incident pulse duration (a), as well as transmitted and reflected spectra when the stack is illuminated with a pulse approximately 250fs in duration (b). We neglect SHG because at normal incidence it is minimized, although anisotropic surface sources might change that. First, we note that transmitted and reflected amplitudes are similar and saturate as a function of pulse duration, as expected, as the signal resolves the resonance. Second, a comparison between Fig.9 (THG from the etalon) and Fig.11 shows that a cavity with a very modest Q can improve THG efficiency by more than two orders of magnitude, despite the fact that the TH is tuned in the metallic range. We note that the widths of the spectra shown in Fig.11(b) depend strictly on the pump spectrum: pulses of increasing duration produce correspondingly narrower TH spectra. This behavior also occurs for a GaAs defect layer surrounded by DBRs when the SH signal is tuned well below the absorption edge, but still in regions of positive dielectric constant. Higher efficiencies are possible by simply adding more periods to the DBRs, as shown for GaAs [44]. These results confirm theoretically the fact that if the pump is tuned to a region of relative transparency, then phase locking triggers a form of electromagnetic transparency at the harmonic wavelengths regardless of material dispersion, with relatively large enhancement of conversion efficiency in cavity environments.

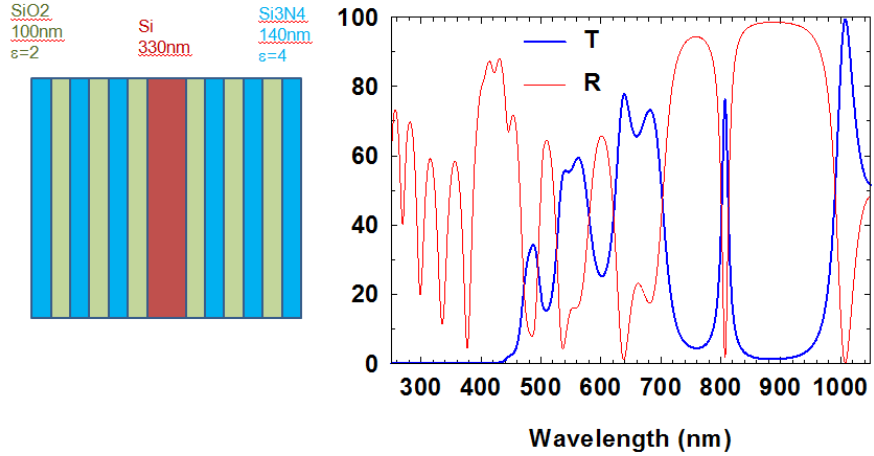

**Fig.10:** Linear transmittance and reflectance of a multilayer that contains a Si defect layer at normal incidence. The bandwidth of the resonance near 800nm is ~20nm, and may be resolved using sub-picosecond incident pulses [44].

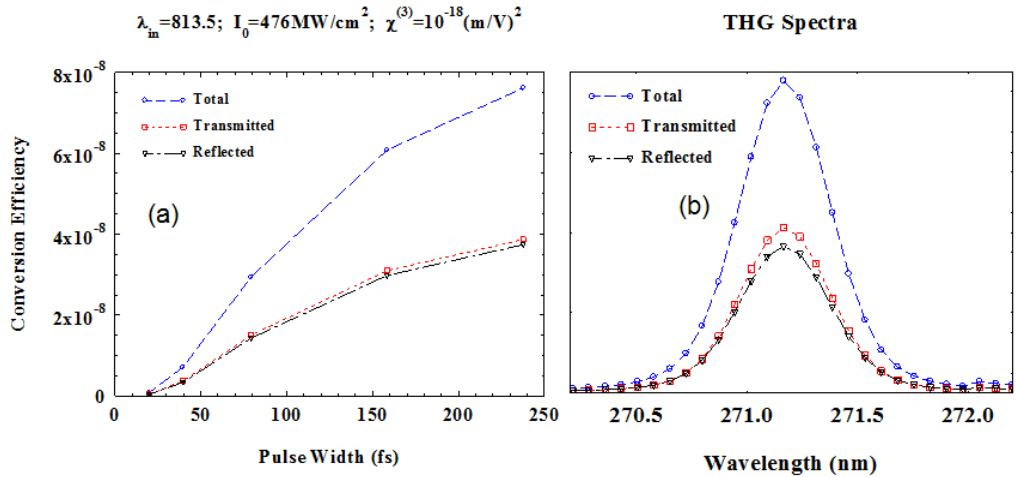

**Fig.11:** (a) Peak transmitted and reflected conversion efficiencies as a function of incident pulse width. The pulses are tuned at resonance near 813nm. THG saturates as longer pulse resolve the resonance. (b) Transmitted and reflected THG spectra when the incident pulse is ~250fs in duration. Even though the cavity Q~40, the total conversion efficiency increases by more than two orders of magnitude compared to the Si etalon.

**Conclusions**

In summary, we have derived a new dynamical model of harmonic generation from centrosymmetric materials that includes surface, magnetic, and electric quadrupole sources. In Silicon, the presence of quadrupoles can remodulate both amplitude and shape of the conversion efficiency of SHG, while THG can be enhanced significantly in cavity environments, even in ranges where the dielectric constant displays metallic behavior. These findings make it possible to rethink the role and functionality of many semiconductors like Silicon or Germanium, perhaps by extending their usefulness as nonlinear media well into the UV range. Finally, it is clear that this approach can be used to model core electron contributions to SHG in metal structures as an extension of the model presented in reference [36]. We have already carried out calculations on structures that contain flat metal surfaces, including the multilayer stacks discussed in reference [36] and find that nonlinear sources associated with the free electron gas continue to dominate. However, similarly to the etalon case that we have discussed, experiments do suggest [43, 45] that additional geometrical features such as surface roughness may turn quadrupoles into a more dominant type of nonlinear source, if the sample is illuminated with multiple beams. Similar effects may occur in metal gratings [37] and it will be interesting to see the predictions this model makes for two-beam excitation geometry.